\documentclass[pra,appendix,superscriptaddress,amsfonts,amssymb,amsmath,twocolumn]{revtex4}

\usepackage{amssymb,amsmath,amsthm,graphicx,amscd,xcolor}

\allowdisplaybreaks

\usepackage{amsfonts}
\usepackage{amssymb}
\usepackage{amsmath}
\usepackage{MnSymbol}
\usepackage{graphicx}
\date{\today}

\begin{document}

\title{Hidden but real: {new} relativistic ``paradox''  exposing the ubiquity of hidden momentum}

\author{Daniel A.\ Turolla \surname{Vanzella}}
\email{vanzella@ifsc.usp.br}
\affiliation{Instituto de F\'\i sica de S\~ao Carlos,
Universidade de S\~ao Paulo, Caixa Postal 369, 13560-970, 
S\~ao Carlos, S\~ao Paulo, Brazil}

\begin{abstract}
The tight connection between mass and energy unveiled by Special Relativity,
summarized by the iconic formula $E = mc^2$, has
revolutionized our understanding of nature and even shaped our political world over the past century through its military application.
It is certainly one of the most exhaustively-tested and well-known equations of modern science.
Although we have become used to its most obvious implication --- mass-energy equivalence
---, it is surprising that one of its subtle --- yet, inevitable --- consequences is
still a matter of confusion: the so-called {\it hidden momentum}.
Often considered as a peculiar feature of specific systems or as an artifact to avoid paradoxal situations,
here we present a {new}
relativistic ``paradox'' which 
exposes the true nature and
ubiquity of
hidden momentum. We also show that  hidden momentum can be forced to reveal itself through observable 
 effects, hopefully putting an end to decades of controversy about its reality.

\end{abstract}

\maketitle

\section{Introduction}
\label{sec:intro}

Einstein's iconic mass-energy relation, $E = mc^2$, is arguably the most famous formula in modern science.
It expresses the equivalence between total mass $m$ and total energy $E$ of a system ($c$ being the speed of light in vacuum),
with wide-ranging consequences:
from the unattainability of the speed of light for massive objects, to particle production in 
high-energy accelerators; from the origin of the energy  of stars  --- less than 
0.1\% of the star's mass, converted into radiation  
over its entire existence ---, to violent bursts of gravitational waves
from merging black holes --- some of them sourced
by several solar masses converted into energy in a fraction of a second. 
Given the importance and generality of  mass-energy equivalence,
it  may strike as a surprise  that one of  its subtle --- but inevitable --- consequences is
still a matter of confusion: the concept of {\it hidden momentum}~\cite{Shockley}
 --- here generalized 
as
the (purely relativistic) part of total
momentum which is not encoded in the motion of the center of mass-energy (CME) of the system. 

In Newtonian mechanics, the total momentum ${\bf P}$ of a {\it closed} mechanical
system --- one which does not exchange {\it matter}  with ``the rest of the universe'' ---
is always given by ${\bf P} = M {\bf V}_{\rm cm}$, where 
$M$ is the total mass of the system and ${\bf V}_{cm}$ is the velocity of its center of mass. This result, 
 known as the center-of-mass theorem, holds true regardless 
 whether the {\it Newtonian} system is subject to external forces or not.
 In contrast to that,
a variety of {\it relativistic} 
systems possessing nonzero total momentum in the rest frame of their CME has 
been identified over the past decades (see, e.g., Refs.~\cite{Shockley,Brevik,Vaidman,Hnizdo92,Hnizdo97,Mansuripur,DV,SB,PS,MK,GH,MansuripurR1,MansuripurR2,
Mansuripur2014,SO}). 
{Here, the term ``relativistic'' does not necessarily mean that large velocities are involved, but 
rather that different inertial-frame descriptions are supposed to be Lorentz covariant  instead of Galilean covariant. 
This covariance constraint  leads to ``unfamiliar'' results (i.e., results 
inexistent in Newtonian mechanics) even in the rest frame 
of the system -- such as non-zero total momentum.}
Such rest-frame momentum has been termed {\it hidden momentum} (HM)~\cite{Shockley}, which now seems to be somewhat
unfortunate because apparently this has
misled many to interpret  its nature as somehow distinct from ``regular'' momentum --- as we argue here, from the 
relativity-theory perspective,
it 
is not. Adding confusion to the story, {\it all} systems in which HM had been identified, until now,
involved  interaction with electromagnetic fields{ --- where it even bears an interesting connection with the 
difference between canonical and kinematic electromagnetic momenta~\cite{SO} ---} and/or moving
 inner parts subject to some
external force field. This masked the generic nature of HM as if it were an exotic feature
--- undesired by some --- of  peculiar 
interaction laws or specific systems.

Here, we present a {new}
relativistic ``paradox'' which shows that  this view is limited and that 
HM is ubiquitous in a relativistic  world {(i.e., a world supposed to be
covariant under Lorentz transformations)}. Moreover, the general definition of HM we propose, freeing its computation from the rest frame of the system,
leads to a  formula {which explicitly shows its relation to asymmetric exchange/flow of 
energy}. 
{Finally, in order to conclusively show that HM is as real as it could be, in the end we discuss an observational
consequence of its existence.}
The present analysis is intended to put an end to decades of confusion about the nature and reality of the so-called 
HM.

The paper is organized as follows. 
In Sec.~\ref{sec:system}, we present the {new} relativistic ``paradox'' involving a heat-conducting bar analyzed from
different inertial-frame perspectives. In order to make the presentation clearer, textbook-level relativistic calculations
which support  statements made in this Section are described in Appendix~\ref{apA}. In Sec.~\ref{sec:notreal},
we put the heat-conducting-bar (HCB) ``paradox'' in context with other known pseudo-paradoxes, stressing their 
origin
in our intuition based on space and time as separate entities rather than in inconsistencies with  known theories.
{We also argue that the HCB ``paradox'' is a close thermal analogue of another relativistic
pseudo-paradox known as ``Mansuripur's paradox''~\cite{Mansuripur}.}
In Sec.~\ref{sec:real}, we argue that HM is just another inevitable consequence of  Einstein's mass-energy
relation $E = mc^2$, 
{obtaining a general expression for HM as dipole moment of energy-exchange rate (Subsec.~\ref{subsec:GeneralHM})
and then applying this general result to solve the HCB ``paradox'' (Subsec.~\ref{subsec:HCBHM})}.
{In Sec.~\ref{sec:realHM}, we show that, contrary to widespread belief expressed in the literature,
the existence of HM in a system can be objectively tested. Finally, in Sec.~\ref{sec:disc}, we present our final comments
and discussion.
It is important to stress that (i)~the HCB pseudo-paradox (Sec.~\ref{sec:system}) and (ii)~the general
deduction of the HM formula (Subsec.~\ref{subsec:GeneralHM}) are independent presentations; the former is discussed only because it evidentiates, in a concrete scenario, the generic nature of HM --- which is the main point of this work.
}

\section{Heat-conducting-bar paradox}
\label{sec:system}

\begin{figure}
\includegraphics[scale=0.27]{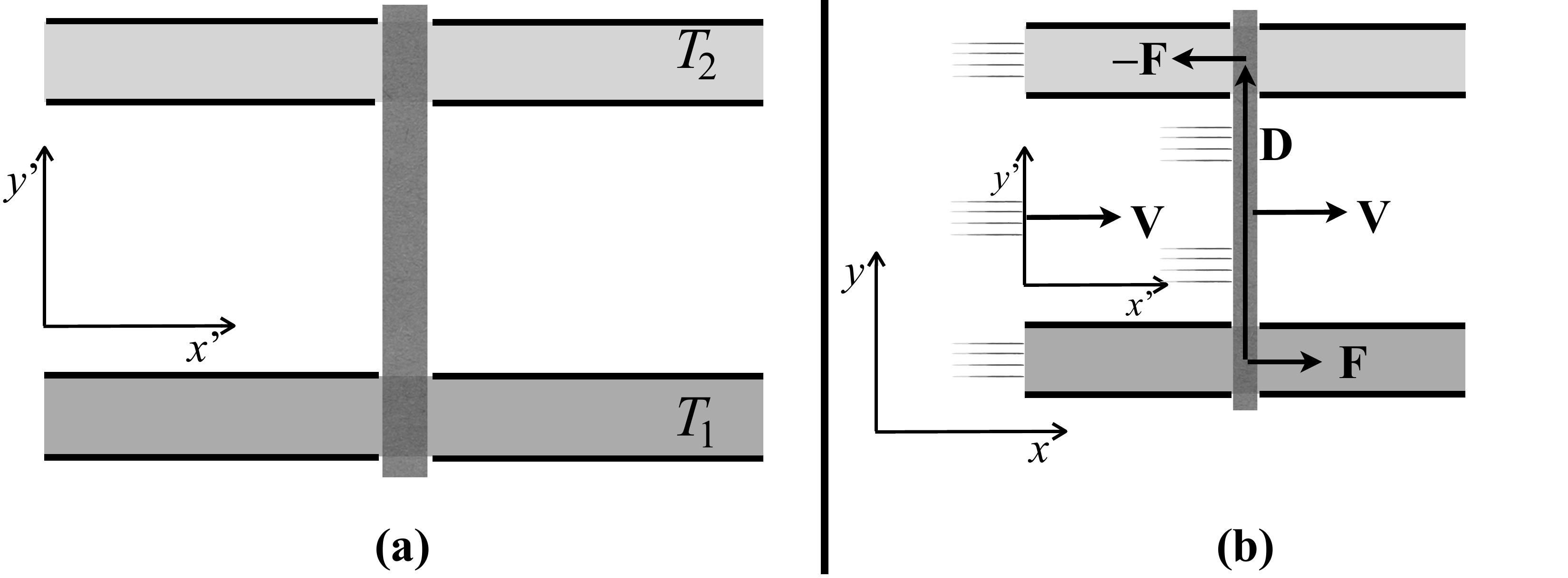}
\caption{Heat-conducting-bar ``paradox.''  (a) A bar stands still in an inertial frame, connecting two 
thermal reservoirs. In the stationary regime, the bar absorbs heat from the thermal reservoir at 
temperature $T_1$, at a rate $W$, and delivers  heat at the same rate to the thermal reservoir
at  temperature $T_2 \leq T_1$. There are no net forces between the bar and the reservoirs. 
(b)~The same situation observed from an inertial frame  w.r.t.\ which the bar moves with velocity ${\bf V}$ perpendicular to itself. 
According to observers static in this latter frame, the reservoirs apply opposite forces 
$\pm {\bf F} =
\pm W{\bf V}/c^2$ on the bar. Therefore, in this frame
there exists a torque ${\bf T} = W ({\bf V}\times {\bf D})/c^2$ on the heat-conducting moving bar, where
${\bf D}$ is the spatial  vector depicted in the figure.}
\label{fig:system}
\end{figure}

Consider the system depicted in Fig.~\ref{fig:system}(a),
composed by a  free bar connecting 
two thermal reservoirs at  temperatures $T_1$ and $T_2$ --- with, say, $T_1 \geq T_2$ ---, at rest 
in an inertial frame. In order to avoid unnecessary
subtleties,
we consider that (i)~the stationary heat-flow regime has been established,
(ii)~thermal 
contact between the bar and each reservoir is symmetric (for instance, through the lateral surface of the bar),
and (iii)~the bar is coated with a thermal insulator all over the parts which are not in contact with the reservoirs. 
Conditions (i) and (ii) ensure that the CME
of the bar stays at rest in the inertial rest frame of the 
reservoirs without the need for any  mechanical constraint; the heat-conducting bar
in the stationary regime is in static mechanical equilibrium.
(Condition (iii) only serves to keep the system simple.)
From the rest-frame perspective, the effect of the reservoirs on the bar is merely exchange of heat, with no net 
forces or torques being applied. Let
 $W>0$ represent the (constant) rate at which  heat is  exchanged between the bar 
 and the reservoirs --- flowing into (respectively,
 out of) 
 the bar from (resp., to)
 the reservoir at temperature $T_1$ (resp., $T_2$). 
(Side note: for any given heat-exchange rate $W$,
 we can consider the temperature difference $T_2-T_1$ to be arbitrarily small by choosing
  bars with arbitrarily large thermal
 conductivities. Therefore, although unnecessary, one can simplify further the setup
 considering  the mass-energy and temperature distributions along the
 bar to be arbitrarily close to homogeneous.)

\begin{figure}
\includegraphics[scale=0.36]{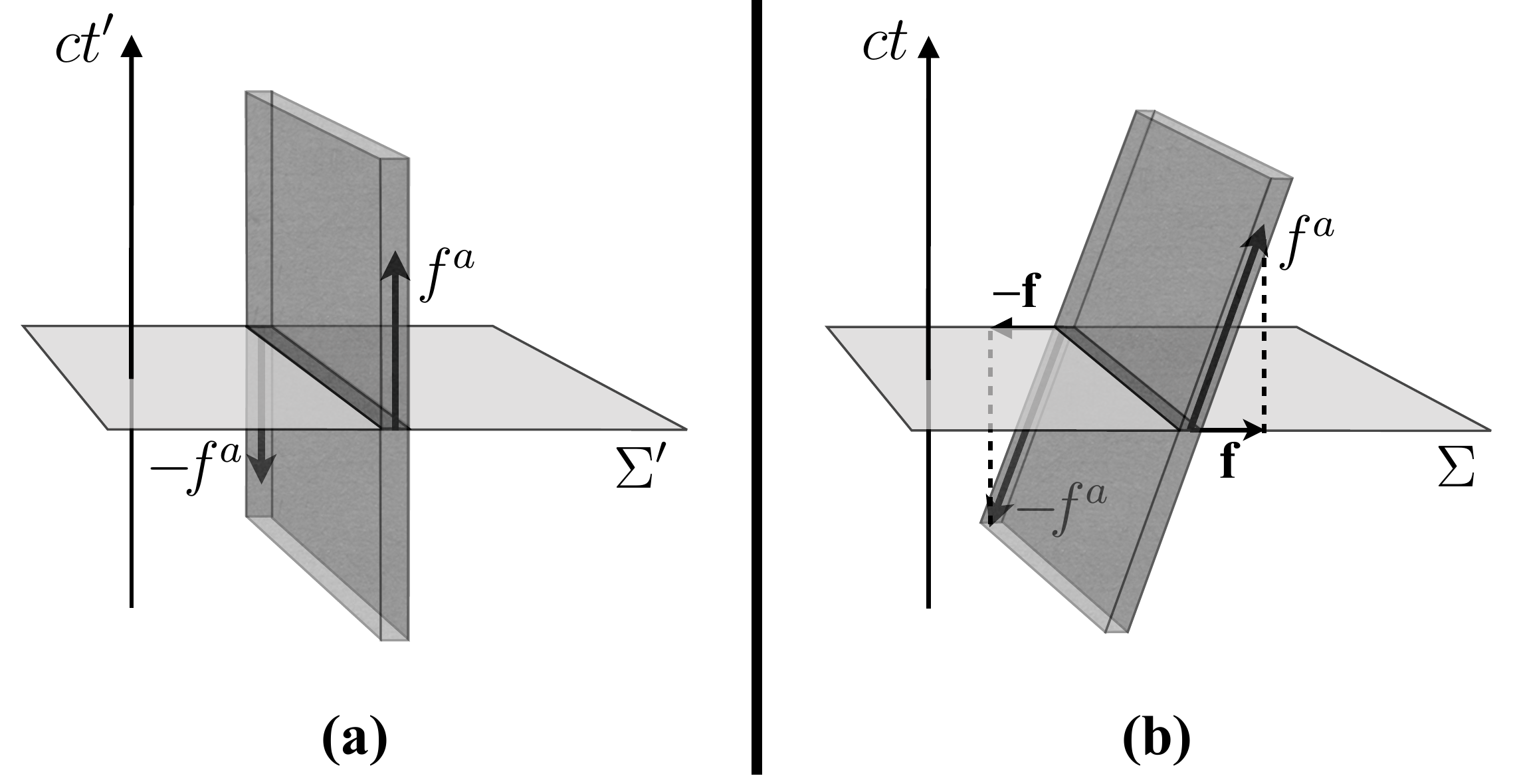}
\caption{Spacetime depiction of the world-volume of the bar and 
the energy-momentum exchange (given by the 4-force densities
$\pm f^a$) between the bar and the reservoirs. (a)~From the rest-frame perspective, $f^a$ has only 
component along the time direction, describing exchange of energy without net spatial forces.
 (b)~From the moving-frame perspective, the {\it same} $f^a$ clearly has nonvanishing spatial projection.
 Therefore, in this frame, there are force densities $\pm{\bf f}$ acting on the bar.}
\label{fig:4force}
\end{figure}

 Now, let us analyze the same setup from the perspective of another inertial frame, with respect to (w.r.t.)~which the bar (and the 
 whole
 system) moves with velocity ${\bf V}$ perpendicular to itself{ --- ``moving frame'' for short}.  
 Although it may sound odd at first,
 it follows directly from Einstein's Special Relativity 
 that,  in  
  this new
  frame,
 the reservoirs apply  opposite  net
 forces $\pm {\bf F} = \pm W {\bf V}/c^2$ at  the moving-bar's ends [see Fig.~\ref{fig:system}(b)].
 The proof of this fact is actually quite simple (a textbook-level exercise) and is 
  explained in detail in Appendix~\ref{apA}. {In essence, due to Lorentz covariance, what is seen in the rest frame
  as a pure exchange of energy (i.e., a $4$-force density $f^a$ with only time component), 
  corresponds to exchange of both energy and
  momentum according to the ``moving frame'' (see Fig.~\ref{fig:4force}, which is the spacetime depiction of the bar in Fig.~\ref{fig:system}).}
 
Once the reader is convinced of  the existence of such forces, 
he/she promptly realizes that they lead to a torque on the heat-conducting moving bar, which
 (neglecting the spatial extension of the thermal contacts) is given by
\begin{eqnarray}
{\bf T} = W ({\bf V}\times {\bf D})/c^2,
\label{THCB}
\end{eqnarray}
 where ${\bf D}$ is the  
 separation vector between the thermal contacts
(see Fig.~\ref{fig:system}); although the opposite forces have no net effect on the total
momentum of the bar as time passes, they do change the bar's angular momentum. 
If we apply our Newtonian
intuition --- as  is customary when arriving at relativistic ``paradoxes'' ---,
this torque w.r.t.~the instantaneous CME position 
should try to rotate the bar.
But this is obviously in conflict with the fact that in the reservoirs' rest frame the bar is in static mechanical equilibrium;
there is absolutely no reason for rotation.
We  have stumbled on a {(new)} relativistic ``paradox.''


\section{No real paradox}
\label{sec:notreal}

Relativistic ``paradoxes'' --- more precisely, situations whose descriptions from different inertial perspectives {\it seem}
paradoxical when compared to each other --- 
are numerous and even serve as teaching tools in relativity. 
Rather than pointing to inconsistencies in fundamental theories, they reveal how our {Newtonian}
 perception  of space and time as separate entities, instead of interwoven in an  absolute four-dimensional spacetime,
 can be deceiving.
Their nature can be loosely  classified as
kinematical --- those which involve only  time-interval and spatial-distance measurements --- and 
dynamical --- those which involve forces. The  twins',  the barn-pole, and the
Bell's spaceship 
``paradoxes'' are well-known textbook samples of the kinematical type --- see, e.g., Ref.~\cite{book} ---, 
whereas the Trouton-Noble~\cite{TN}, the  right-angle-lever~\cite{rightanglelever}, and 
the submarine~\cite{submarineS,submarineM,submarineK,submarineB}
 ``paradoxes'' are representative of the dynamical type. 
 The heat-conducting-bar (HCB) ``paradox'' presented above clearly fits into this latter class.
 Contrary to 
kinematical ``paradoxes,'' the dynamical ones are rarely addressed in relativity textbooks and introductory courses. This may explain why many of them are 
unknown to nonrelativists or, when known, concepts involved in their resolution are 
seen with suspicion. 

\begin{figure}[h]
\includegraphics[scale=0.25]{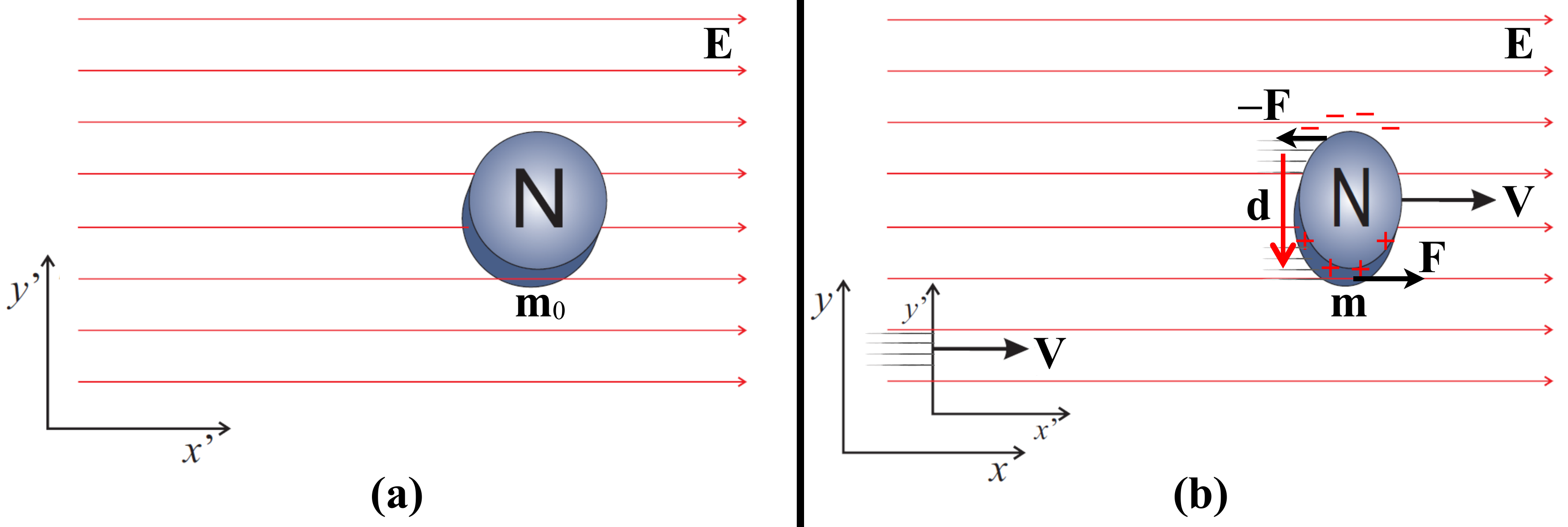}
\caption{Mansuripur's  ``paradox.'' (a) A magnet stands still in an inertial frame, 
with magnetic dipole moment ${\bf m}_0$ perpendicular to a uniform electric field
${\bf E}$. (b)~The same situation observed from an inertial frame w.r.t.\ which the magnet moves with 
velocity ${\bf V}$ parallel to the electric field. Now, the magnet also carries an electric
dipole moment ${\bf d}$, upon which the electric field exerts a torque.}
\label{fig:magnet}
\end{figure}

In 2012, 
M.\ Mansuripur~\cite{Mansuripur} analyzed in detail an ingenious dynamical ``paradox''  --- previously discussed in 
Ref.~\cite{Namias} --- 
which, in a simplified but equivalent version, can be
realized by a neutral 
magnet at rest in an inertial frame, where there exists a uniform (external) electric field ${\bf E}$ perpendicular to the magnet's magnetic dipole moment ${\bf m}_0$ (see Fig.~\ref{fig:magnet}).
In the magnet's rest frame [Fig.~\ref{fig:magnet}(a)], the magnet ``seems'' oblivious to the presence of the electric 
field --- apart from
induced polarization, which can be made negligible.
However, looking at the same system from another  inertial frame, w.r.t.\ which the magnet moves
with velocity ${\bf V}$ 
 along the electric-field's direction [Fig.~\ref{fig:magnet}(b)], 
the magnet now also bears an electric dipole moment  ${\bf d} = {\bf V}\times {\bf m}_0/{c}^2$ --- since 
${\bf m}_0$ is ultimately due to electric currents, not pairs of magnetic monopoles~\cite{Gr92}.
 Thus, according to this inertial frame, there must exist a torque 
${\bf T} = {\bf d}\times {\bf E} = ({\bf V}\cdot {\bf E})\,{\bf m}_0/c^2 $
acting on the magnet, which would supposedly make it spin --- in gross contradiction with  the fact that in its inertial rest frame the magnet stands still. Mansuripur
concludes that this contradiction is an ``incontrovertible theoretical evidence of the
incompatibility of the Lorentz law [of force] with the fundamental tenets of special relativity''~\cite{Mansuripur}.

\begin{figure}[h]
\includegraphics[scale=0.35]{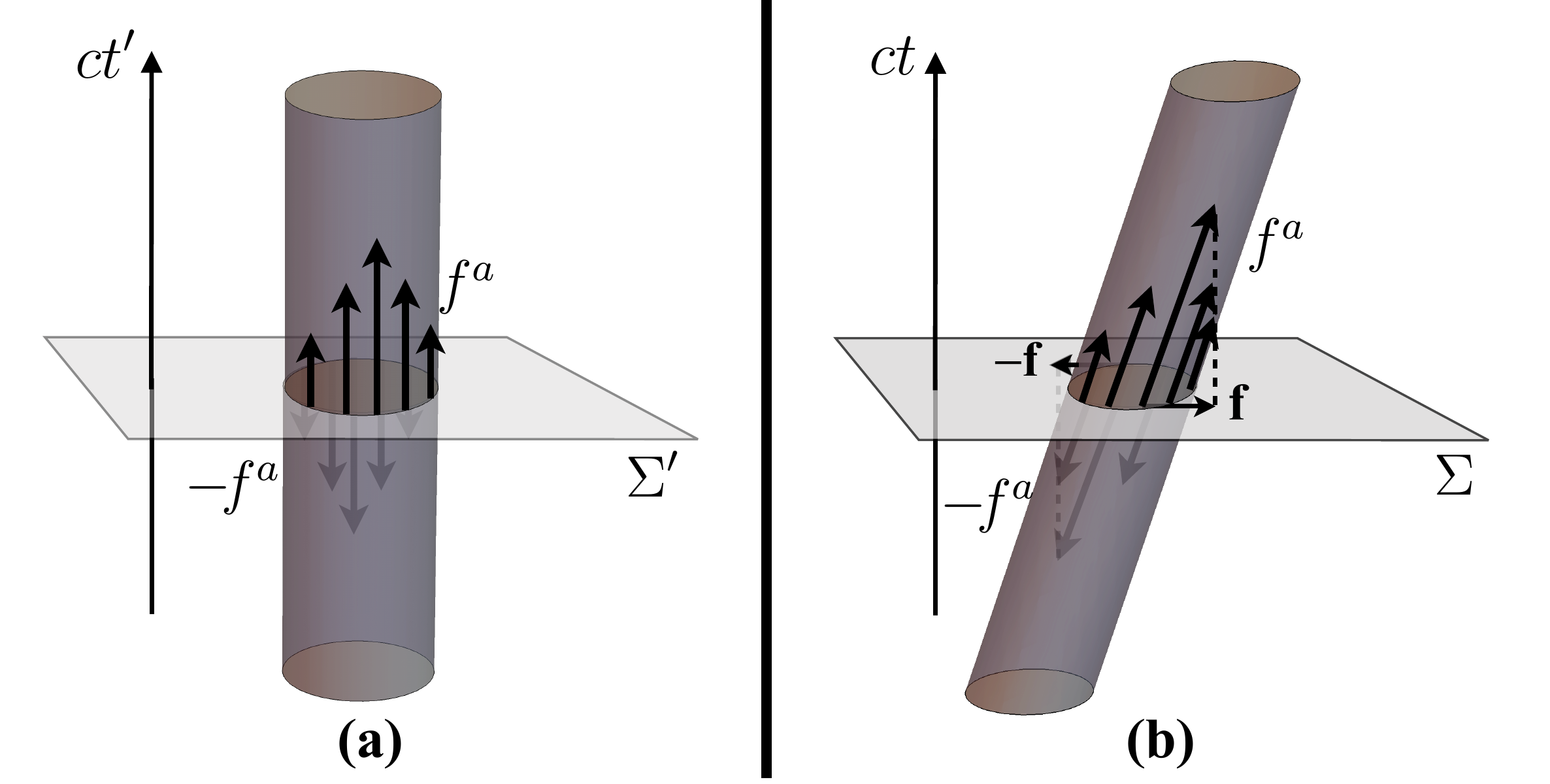}
\caption{{Four-dimensional representation of (the cross section $z=0$ of) the system
depicted in Fig.~\ref{fig:magnet}: (a) Privileging the magnet's rest frame;
(b) Privileging the
``moving frame.''
The Lorentz 4-force density 
is future-directed ($f^a$) where the electric field favors the current density (${\bf j}\cdot {\bf E}>0$)
and past-directed otherwise 
(${\bf j}\cdot {\bf E}<0$).
Note that it has null projection on $\Sigma^\prime$ --- therefore, no forces according to the magnet's rest frame ---
 while being nonzero
and circulating  on ${\Sigma}$ --- therefore, applying a torque on the magnet according to the ``moving frame.''}}
\label{fig:4dmagnet}
\end{figure}

The HCB ``paradox'' presented earlier is a close
 thermal analogue of Man\-su\-ri\-pur's, with the thermal
reservoirs playing the role of the external electric field
{and} the heat-conducting bar substituting the magnet. Less 
obvious is the analogue, in Mansuripur's setup, of
the heat exchange rate $W$ {between the bar and the reservoirs.} 
{But recalling that magnetism in materials is ultimately due to current densities
${\bf j}$ (even if quantum mechanical in nature), one concludes that 
the magnet in its rest frame does exchange energy with
the external electric field at a}
 rate, per volume, ${\bf j}\cdot {\bf E}${: the magnet predominantly absorbs 
(resp., delivers)
energy from (resp., to) the external field where ${\bf j}\cdot {\bf E}>0$ (resp., ${\bf j}\cdot {\bf E}<0$),
leading to a net flow of energy
across the magnet. This completes the analogy between the HCB  ``paradox'' and Mansuripur's
(compare Fig.~\ref{fig:4dmagnet} with Fig.~\ref{fig:4force}).} 
{Notwithstanding,
there is one important difference:} 
in the thermal analogue, there is no specific ``law of force'' to blame for the apparent contradiction between different inertial-frame
descriptions; the torque on the bar seen from the ``moving-frame'' perspective is 
{{\it enforced} by Lorentz covariance and, particularly, by} $E = mc^2$.
Certainly, no one would hold that 
Einstein's mass-energy relation is ``incompatible with the fundamental tenets of special relativity.''
Therefore, there is no logical reason for taking this stand regarding the Lorentz force.

\section{{Mass-energy equivalence and Hidden Momentum}}
\label{sec:real}

Relativistic thermodynamics has its own history of subtleties and
  controversies.  The most emblematic of them is probably the question of how 
 temperature transforms from one inertial frame to another. 
 It took about 90 years for this to be recognized as
 an ill-posed question --- hence, the conflicting answers given during this period  (see Refs.~\cite{Sandro,LM} and 
 references therein).
 Fortunately, none of these subtleties --- not even temperature transformation --- 
 concerns us; the purpose of thermal reservoirs in the setup of 
 Fig.~\ref{fig:system} is only to guarantee
 an eventual stationary situation in the rest frame of the system.
 
 As mentioned earlier,
several systems with nonvanishing total momentum in their rest frames have
been found and discussed in the literature (see, e.g., Refs.~\cite{Shockley,Brevik,Vaidman,Hnizdo92,Hnizdo97}) ---
including Mansuripur's setup~\cite{DV,SB,PS,MK,GH,SO}. 
All such systems
involved  interaction with electromagnetic fields and/or moving
 inner parts subject to some
external force field, which led many to view it as a feature of peculiar
interaction laws or systems.
Mansuripur, for instance, 
considered HM to be an  {\it ad hoc} 
addition to  materials interacting with electromagnetic fields, with no justification other than artificially 
avoiding paradoxal situations~\cite{MansuripurR1,MansuripurR2,Mansuripur2014}.
A better law of electromagnetic force, he reasoned, should be one which  leads to no torque on the moving magnet in
an electric field --- hence, doing away with HM. 
In this sense, the HCB ``paradox'' we discuss in this work is unique, 
for it does not depend on the inner details of the system 
(the bar and the 
heat/energy flow) and of the interaction with ``the rest of the universe'' (the thermal/energy reservoirs).
 
 The resolution of the HCB ``paradox'' 
 --- as well as Mansuripur's --- 
 consists in taking mass-energy equivalence to its ultimate consequences.
 As heat (i.e., energy) flows through the bar, it contributes to momentum in very much the same way as would a flow of
 matter. 
  In fact, distinguishing contributions to the total momentum
coming from ``different forms'' of energy flows is quite contrary to the spirit
of relativity theory.
 Therefore, the total momentum
 of the bar in its {\it rest} frame [Fig.~\ref{fig:system}(a)] does not vanish --- a purely relativistic effect.
For the same reason,
according to the inertial frame w.r.t.\ which the bar moves with velocity ${\bf V}$ perpendicular to itself
[Fig.~\ref{fig:system}(b)], there is a momentum contribution along the bar. Consequently,  the bar's total momentum
${\bf P}$ and  the CME velocity ${\bf V}$ are misaligned; and
dragging momentum ${\bf P}$ along a spatial direction 
which is not
aligned to it inevitably leads to
 a time-varying angular momentum ${\bf L}$ (with $d{\bf L}/dt= {\bf V}\times {\bf P}$)
 and, therefore, {\it demands}  a torque --- which, as we shall see below, is
 precisely the one supplied by the thermal reservoirs in the moving frame. 

\subsection{{General treatment: hidden momentum as dipole moment of energy-exchange rate}}
\label{subsec:GeneralHM}

{Mass-energy equivalence has a very straightforward consequence which, nonetheless, is 
overlooked
when arriving at dynamical relativistic ``paradoxes:''  the distinction between {\it closed} and 
{\it isolated} systems, as
usually  made in Newtonian (i.e., Galilean-covariant) physics 
(including nonrelativistic thermodynamics), has no absolute 
meaning in relativistic physics. While in Newtonian physics a system can exchange energy without exchanging mass
(i.e., it can be non-isolated but closed),
this is obviously impossible if mass and energy are the same physical quantity. 
In this sense, the {\it relativistic} dynamics of the heat-conducting bar depicted in Fig.~\ref{fig:system} --- which, in 
the Newtonian context, is a closed system for which the center-of-mass theorem would apply --- is essentially
the same as that of a 
pipe segment carrying a steady fluid/particle current, with fluid/particles entering the system at one end of the pipe
 and leaving at the other --- which, in the Newtonian context, is an open system, for which the center-of-mass
 theorem does {\it not} apply.
  No 
 one would object that the pipe segment carrying a particle/fluid current  possesses  nonzero momentum in the
 rest frame of its center of mass. 
 Distinguishing this momentum from  the one carried by the heat-conducting bar --- or, for that matter, from the one carried 
by the magnet in Mansuripur's setup --- is solely  motivated by our Newtonian view of the world.
}

{Although 
distinguishing  HM from ``regular'' momentum is artificial
as far as relativistic dynamics is concerned, 
it is useful, in order to demystify it further, to  
pin down the elements which compel our Newtonian intuition to make such a distinction.
Basically, all instances of HM involve systems where there is a clear
  notion of a {\it velocity field} ${\bf v}$ of their ``constituents'' 
 --- usually taken  to be   particles  or fluid elements --- and an associated nonnegative (not identically null) 
 {\it  number density}
 $n$ which, together, satisfy the continuity equation 
 $ \partial_t n + \nabla \cdot (n{\bf v}) = 0$ in {\it all} inertial frames. 
 Put  in spacetime language: these systems in which
 HM can be identified posses, associated to their constituents,  a  time-like, future-directed 4-vector field $n^a$ --- the
 4-current number density, whose components in inertial Cartesian coordinates read $n^\mu = (n c,n{\bf v})$
 --- satisfying the tensorial equation $\nabla_a n^a = 0$. The key point is that the existence of such a 4-current 
  number density allows us to extend to the relativistic context, in a consistent manner, the notion  of ``closed systems:''
}
\\
\begin{description}
\item {
 {\it Definition:} A system
will be said to be {\it closed} if  one of the following holds:}
\begin{description}
\item {(a) The system is {\it isolated} --- i.e., its {\it extended}  (see below) stress-energy-momentum tensor
$\bar{T}^{ab}$
satisfies $\nabla_a\bar{T}^{ab} \equiv 0$ (and goes to zero sufficiently fast at spatial infinity so that any flux vanishes);}
\item  {(b) The system possesses a ``natural'' notion of 4-current number density $n^a$ (as defined above)
whose {\it extension} $\bar{n}^a$ (see below)
satisfies $\nabla_a \bar{n}^{a} \equiv 0$ (and goes to zero sufficiently fast at spatial infinity so that any flux vanishes);}
\item  {(c) The system itself is a collection of closed systems as defined in the previous items.}
\end{description}
{
(Given a tensor field ${\cal T}^{ab...}_{cd...}$ with support $supp({\cal T})$, its extension
$\bar{{\cal T}}^{ab...}_{cd...}$ is the tensor field defined over the {\it whole} spacetime which
coincides with ${\cal T}^{ab...}_{cd...}$ in $supp({\cal T})$ and is zero otherwise. This is a mere
technicality, useful when treating systems which are, themselves, part of larger  ones. 
Due to possible discontinuities in $\bar{\cal T}^{ab...}_{cd...}$, the equations above are to be taken in the distributional sense.)}
\end{description}
 
 {The definition above clearly recovers, in the Newtonian regime, the notion of closed systems as those which do
 not exchange matter/mass, for in this case the mass density itself satisfies the continuity equation and can be
 taken to be $n$ up to a multiplicative constant. This fact will be used later when we restrict attention to closed systems.}

Let ${\cal S}$ be a system {with}
{stress-}energy-momentum tensor whose components in  inertial
Cartesian coordinates $\{({c}t,{\bf x})\}$ are given by $T_{\cal S}^{\mu \nu}$. We assume that 
at each instant $t$, $T_{\cal S}^{\mu \nu }${ goes to zero sufficiently fast at spatial infinity so that 
the manipulations and integrals which follow
below are well defined}.
The center of mass-energy (CME) of ${\cal S}$ is given by
\begin{eqnarray}
{\bf X}_{\rm cme} := \frac{1}{Mc^2}\int d^3x \;\bar{T}_{\cal S}^{00} \;{\bf x},
\label{Xcm}
\end{eqnarray}
where $M = \int d^3x \;\bar{T}_{\cal S}^{00}/c^2$ is the (possibly time-dependent) total mass
of the system. {(All  integrations are carried over the entire spatial sections $t = {\rm constant}$; hence the use
of $\bar{T}^{\mu \nu}_{\cal S}$ instead of ${T^{\mu \nu}_{\cal S}}$.)}
Multiplying Eq.~(\ref{Xcm}) by $M$ and taking the time derivative, we 
get:
\begin{eqnarray}
M {\bf V}_{\rm cme}&=&-\frac{dM}{dt}
{\bf X}_{\rm cme} + \frac{1}{c}\int d^3x\;
\partial_0 \bar{T}_{\cal S}^{00} \; {\bf x}  
\nonumber \\
&=&\frac{1}{c} \int d^3x\;
\partial_0 \bar{T}_{\cal S}^{00} \; ({\bf x}-{\bf X}_{\rm cme}),
\label{manip}
\end{eqnarray}
where ${\bf V}_{\rm cme} :=d{\bf X}_{\rm cme}/dt$ is the velocity of the CME of ${\cal S}$.

The fact that system ${\cal S}$ {is not necessarily isolated} means that
 $\partial_\mu \bar{T}_{\cal S}^{\mu \nu}=f^\nu$, 
where $f^{\mu}$ is the 4-force density acting on ${\cal S}$ --- in particular, $f^0$ is related to  
energy exchange rate
$W$ 
 through $d^3x \,f^0 = dW/c$. Substituting
$\partial_0 \bar{T}_{\cal S}^{00}= f^{0}-\partial_j \bar{T}_{\cal S}^{j0}$  into Eq.~(\ref{manip}) leads to:
\begin{widetext}
\begin{eqnarray}
M {\bf V}_{\rm cme}-\frac{1}{c} \int d^3x\;
f^0 \; ({\bf x}-{\bf X}_{\rm cme})
\!\!\!&=&\!\!\!-\frac{1}{c}\int d^3x\;
\partial_j \bar{T}_{\cal S}^{j0} \; ({\bf x}-{\bf X}_{\rm cme})
\nonumber \\
&=&\!\!\!-\frac{1}{c}\int d^3x\;
\partial_j [\bar{T}_{\cal S}^{j0} \; ({\bf x}-{\bf X}_{\rm cme})] 
+\frac{1}{c}
\int d^3x\;
\bar{T}_{\cal S}^{j0} \; \partial_j{\bf x}
\nonumber \\
&=&
\int d^3x\;
{\bf p}={\bf P} ,
\label{manip2}
\end{eqnarray}
\end{widetext}
where $({\bf p})^j = \bar{T}_{\cal S}^{j0}/c$ are the components of the momentum 
density of the system.

{So far, very little has been imposed on the system ${\cal S}$. In fact, the only assumptions are that
the integrals above converge and that the surface term coming from the first integral in the right-hand side 
(r.h.s.)~of the second line of Eq.~(\ref{manip2}) vanishes at spatial infinity. But now, we restrict attention to closed systems, as defined earlier. The reason is that for closed systems,} the second term in the left-hand side {(l.h.s.)}~of Eq.~(\ref{manip2}) is 
purely relativistic, since {in the Newtonian regime $n$ is proportional to mass density and, therefore, 
$f^0/c \propto \nabla_a \bar{n}^a \equiv 0$}. 
This expresses the well-known fact  that in Newtonian mechanics, the total momentum of an arbitrary {closed} system 
(isolated or not) is completely
encoded in the motion of its center of mass 
and its total mass{ --- the center-of-mass theorem}.
In relativity theory, 
on the other hand, {we see that} asymmetric (w.r.t.~the CME) exchanges of energy
between 
${\cal S}$ and ``the rest of the universe'' contribute to the total momentum of the system; now, ${\bf P}$ 
cannot be assessed only by keeping track of the system's  mass-energy distribution. This  motivates us
 to define the  ``hidden'' part of the total momentum of the system as 
${\bf P}_{\rm h}:= {\bf P} - M {\bf V}_{\rm cme}$, which
can then be calculated by:
\begin{eqnarray}
{\bf P}_{\rm h}=-\frac{1}{c}
\int d^3x\;
f^0 \; ({\bf x}-{\bf X}_{\rm cme}) = -\frac{1}{c^2}\int dW ({\bf x}-{\bf X}_{\rm cme}).
\label{Phfinal}
\end{eqnarray}
In words: the hidden momentum of a {closed} system is given by (minus $1/c^2$ times) 
the dipole moment (w.r.t.~${\bf X}_{\rm cme}$)
of {its} energy-exchange rate. 
{Notice that our
definition not only frees HM from being  identified only in 
the rest frame of the system (where ${\bf V}_{\rm cme}= {\bf 0}$), but  also shows that HM does not
depend 
on the inner details of the system; it does not depend on the nature 
of $T^{ab}_{\cal S}$ (electromagnetic, thermal, mechanical, etc.), but only on how it fails to be
conserved.}\footnote{After acceptance of this manuscript for publication, 
Ref.~\cite{Boyer2014} came to our attention, where  ``internal momentum,'' defined in a {\it similar} 
manner as HM here, is  
proposed to substitute the latter. 
We essentially agree with the key points of Ref.~\cite{Boyer2014}, whose attempt to demystify HM is 
aligned with ours. Notwithstanding, 
Ref.~\cite{Boyer2014} still limits its discussion to an electrodynamic
system, restricting energy exchanges  to mechanical work: $f^0 = {\bf f}\cdot {\bf v}/c$. 
In particular, its definition of ``internal momentum'' [following from its Eq.~(5)] would vanish for the scenario depicted in
Fig.~\ref{fig:system}(a), in conflict with both our Eq.~(\ref{PhHCB}) and   the observable effect
predicted in Sec.~\ref{sec:realHM}.
 The HCB ``paradox,''
the precise characterization of systems which  bear HM, and the
observational consequence of HM which we discuss in what follows 
constitute important new ingredients to showing that
HM (or ``internal momentum'') is a 
{\it generic} relativistic feature --- as we had already claimed in Ref.~\cite{DV}.}

\subsection{{Hidden momentum in a heat-conducting bar}}
\label{subsec:HCBHM}

{Applying the definition given in Eq.~(\ref{Phfinal}), or its discrete version}
\begin{eqnarray}
{\bf P}_{\rm h} = -\frac{1}{c^2}\sum_{j}  ({\bf x}_j -{\bf X}_{\rm cme}) W_j,
\label{PhW}
\end{eqnarray}
{to the system depicted in Fig.~\ref{fig:system}, we promptly obtain:}
\begin{eqnarray}
{\bf P}_{\rm h} = W {\bf D}/c^2.
\label{PhHCB}
\end{eqnarray}
{Therefore, in the situation depicted in Fig.~\ref{fig:system}(b), dragging the total momentum
${\bf P} = M{\bf V}+{\bf P}_{\rm h}$
of the bar at a constant velocity ${\bf V}$ leads to an angular momentum ${\bf L}$ which changes
at a rate}
$$
\frac{d{\bf L}}{dt} = {\bf V}\times {\bf P} = {\bf V}\times {\bf P}_{\rm h} = W\,{\bf V}\times {\bf D}/c^2.
$$
{Comparing this result  
with the torque  given in Eq.~(\ref{THCB}), provided by the forces $\pm{\bf F}$, we see that everything
fits perfectly: the torque provided by the forces seen from the moving frame is exactly the one needed
to keep the spinless bar in uniform motion.}

{Obviously, Eq.~(\ref{Phfinal}) can also be applied to the magnet depicted in Fig.~\ref{fig:magnet}~\cite{DV}.
Recalling that ${\bf j} = \nabla\times {\bf M}$, where ${\bf M}$ is the magnet's magnetization, we have:}
\begin{eqnarray}
{\bf P}_{\rm h}&=&-\frac{1}{c^2}
\int d^3x\;
({\bf j}\cdot {\bf E}) \; ({\bf x}-{\bf X}_{\rm cme})\nonumber \\
&=&-\frac{1}{c^2}\int d^3x\;[\nabla\cdot({\bf M}\times {\bf E})] \; ({\bf x}-{\bf X}_{\rm cme})\nonumber \\
&=&\frac{1}{c^2}\int d^3x\;({\bf M}\times {\bf E}) = \frac{1}{c^2}\left(\int d^3x\;{\bf M}\right)\times {\bf E} \nonumber \\
&=& \frac{1}{c^2}{\bf m}_0\times {\bf E}.
\label{PhMagnet}
\end{eqnarray}
{Therefore, in situation depicted in Fig.~\ref{fig:magnet}(b), the total angular momentum ${\bf L}$ of the magnet
changes at a rate:}
$$
\frac{d{\bf L}}{dt} = {\bf V}\times {\bf P} = {\bf V}\times {\bf P}_{\rm h} = ({\bf V}\cdot {\bf E}) {\bf m}_0/c^2,
$$
{which matches exactly the torque applied on the magnet according to the ``moving frame''
--- see   Sec.~\ref{sec:notreal}.}


\section{{Hidden but real... and observable}}
\label{sec:realHM}

{By now, we hope we have convinced the reader that HM, far from being a peculiar property of specific systems
or interaction laws, is simply a legitimate relativistic contribution to total momentum coming from energy flows in a closed,
non-isolated system --- a distinction motivated solely by our Newtonian intuition. 
In this sense, not only it is ubiquitous in a relativistic world, but also it is 
as real as any other form of
momentum. In fact, we present below a final conclusive evidence which corroborates this view: 
we show that HM can be converted into
``regular'' momentum and, as such, its existence has observable consequences.}

{Consider the same system depicted in Fig.~\ref{fig:system}(a) in stationary regime of heat flow. Suppose, now, 
that the thermal contacts with both reservoirs are suddenly interrupted simultaneously in the bar's frame (e.g., by some clever preprogrammed mechanism attached to the
bar itself which shields the contacts with some thermal insulator). Since this interruption can be performed without any external forces applied on the bar, its total momentum will  not change in the process. However, once the thermal contacts are interrupted, the bar is isolated and, as such, the center-of-mass theorem must hold. This implies that
the bar's initial total momentum, which was completely ``hidden,'' now has to manifest itself as motion of the bar's CME:
}
$$
{\bf V}_{\rm cme} = \frac{{\bf P}_{\rm h}}{M} = \frac{W {\bf D}}{Mc^2},
$$
{where $M$ is the mass of the bar in situation depicted in Fig.~\ref{fig:system}(a). (Note that if, in preparing 
the setup represented in Fig.~\ref{fig:system}(a), we had not held the bar fixed until the stationary heat-flow regime
had been established, the bar might have acquired a velocity in the opposite direction 
in order to keep its 
total momentum zero --- in case thermal contacts 
are established
with no net external  forces acting on  the bar in its rest frame. This is the reason for assumption (i) made in Sec.~\ref{sec:system}.) Although this
velocity is probably too small in  realistic situations, it constitutes evidence that HM is not some imaginary concept 
with no objective existence. (Obviously, HM contained in the ``rest of the universe'' will also be converted into
motion of the CME of this latter, but since the ``rest-of-the-universe'' mass is supposedly much larger
than that of the system of interest, such effect may be completely neglected.)}

{The same conclusion holds for the magnet represented in Fig.~\ref{fig:magnet}: {\it if} the electric field is 
removed  without exerting net forces on the magnet --- which may be difficult due to possibly inhomogeneous 
magnetic fields generated in the process ---, the magnet would acquire a velocity
}
$$
{\bf V} = \frac{{\bf P}_{\rm h}}{M}= \frac{{\bf m}_0\times {\bf E}}{Mc^2},
$$
{with $M$ being the magnet's mass. 
This may be seen as somewhat similar to the Richardson--Einstein--de Haas 
effect~\cite{Richardson,EdH},
but now for linear  instead of angular velocity/momentum. 
(In this hypothetical scenario, it seems reasonable to conjecture that the opposite momentum  of the ``rest of the 
universe'' may become manifest through emission of electromagnetic waves.)
Since both $M$ and ${\bf m}_0$, in ideal situations, scale with the magnet's
volume, we can make an order-of-magnitude estimation for $V: = \|{\bf V}\|$ using Bohr's magneton, 
$m_0:=\|{\bf m}_0\|\sim\mu_B\sim 10^{-13}$~GeV$/$T, and the proton's mass, $M\sim M_p\sim 1$~GeV$/c^2$:
$V\sim [E/(1~{\rm kV}/{\rm m})]\times 10^{-1}$~nm$/$s, where $E := \|{\bf E}\|$. As expected, the typical velocities
for real magnets in realistic external electric fields are extremely small. However, one might try to amplify this 
effect by using
systems in which $M$ scales with size slower than does $m_0$ --- as in solenoids, in the macroscopic scenario, or
Rydberg atoms, in the atomic realm. }

\section{Discussion}
\label{sec:disc}

Although  Mansuripur's speculation on alternative laws of  electromagnetic force is a valid inquiry --- 
which can only be definitely  settled by experiments ---, the generic 
nature of the HCB ``paradox'' ---  with electromagnetism and moving inner parts playing no explicit essential role ---
shows that the existence  of torques acting on spinless, 
uniformly-moving objects is a ubiquitous feature of relativistic dynamics. As stated
earlier,
this torque (${\bf T} = {\bf V}_{\rm cme}\times {\bf P}$) is responsible for translating 
the CME of
the system (with velocity ${\bf V}_{\rm cme}$) along a direction which is not aligned to its total momentum (${\bf P}$)
--- which exposes the existence of HM. 
{The} generalized definition of HM as
\begin{eqnarray}
{\bf P}_{\rm h} := {\bf P} - M{\bf V}_{\rm cme},\nonumber 
\end{eqnarray}
{proposed in Subsec.~\ref{subsec:GeneralHM}} --- which makes sense not only in the rest
frame of the system (where ${\bf V}_{\rm cme} = {\bf 0}$) ---, 
{led} to a formula relating
HM  with  asymmetric (w.r.t.~the system's CME)
exchange of energy  with  ``the rest of the universe:''
\begin{eqnarray}
{\bf P}_{\rm h} = -\frac{1}{c^2}\sum_{j}  ({\bf x}_j -{\bf X}_{\rm cme}) W_j,
\label{PhWconc}
\end{eqnarray}
where ${\bf X}_{\rm cme}$ is the CME position and ${\bf x}_j$ is the position  where energy exchange occurs 
at a rate $W_j$ ($W_j>0$, if energy enters the system; $W_j<0$, if energy leaves the system).
 The interpretation is simple: this asymmetry leads to 
energy flows in the system which, regardless their 
nature, contribute to momentum in very much  the same way as  do 
matter flows  --- thanks to mass-energy 
equivalence. As stressed earlier, distinguishing contributions to the total momentum
coming from ``different forms'' of energy flows is quite contrary to the spirit
of relativity theory --- reason why
a covariant, observer-independent definition of HM does not (and cannot) exist. Notwithstanding,
although artificial and not strictly necessary, thinking in terms of HM may help our Newtonian intuition
to spot effects which might pass unnoticed otherwise --- as the one discussed in Sec.~\ref{sec:realHM}.

Obviously, only experiments can decide on the correctness of candidate  laws of Nature.
{For instance, whether or not HM is in fact present in a magnet subject to an external electric field
strongly depends on the ultimate 
origin of its magnetic
dipole moment~\cite{SO}. Here, in order to arrive at the well-known  Eq.~(\ref{PhMagnet}), we have adopted the 
standard view that any magnetic moment is ultimately
due to electric currents, even if of a quantum nature. Notwithstanding, 
 Eq.~(\ref{PhWconc}) is equally valid whatever is the microscopic modeling of magnetic dipoles.} 
{Be it as it may, the point is that}
aiming at substituting a law of force
solely  on the basis that it leads to HM is a misguided effort. As made explicit by the 
HCB ``paradox'' and Eq.~(\ref{PhWconc}), HM is simply an inevitable consequence
of $E = mc^2$ when seen from an arbitrary inertial frame. {Moreover, its existence can, 
in principle, be tested by forcing it
to reveal itself as motion of the CME of the system ---
as shown in Sec.~\ref{sec:realHM}. Looking at this the other way around, 
measuring the amount of HM in atomic-size ``magnets'' subject to  external fields may end up being useful for 
investigating or confirming
the
ultimate nature of elementary magnetic moments.
}


\acknowledgments
The author thanks Kirk T.\ McDonald, Vladimir Hnizdo, David J.\ Griffiths, and  Massud Mansuripur
  for discussions which  motivated{ --- not necessarily implying 
  agreement with ---} this work. In particular, we thank 
  Kirk T.\ McDonald for calling attention to Ref.~\cite{Boyer2014}. The author also thanks Barnab\'as Deme
  for discussions regarding the similarity between the effect discussed in Sec.~\ref{sec:realHM}
  and the 
 Richardson--Einstein--de Haas effect.

\appendix

\section{Energy-momentum transfer in  inelastic collisions and force exchange with thermal reservoirs}
\label{apA}

\begin{figure}
\includegraphics[scale=0.25]{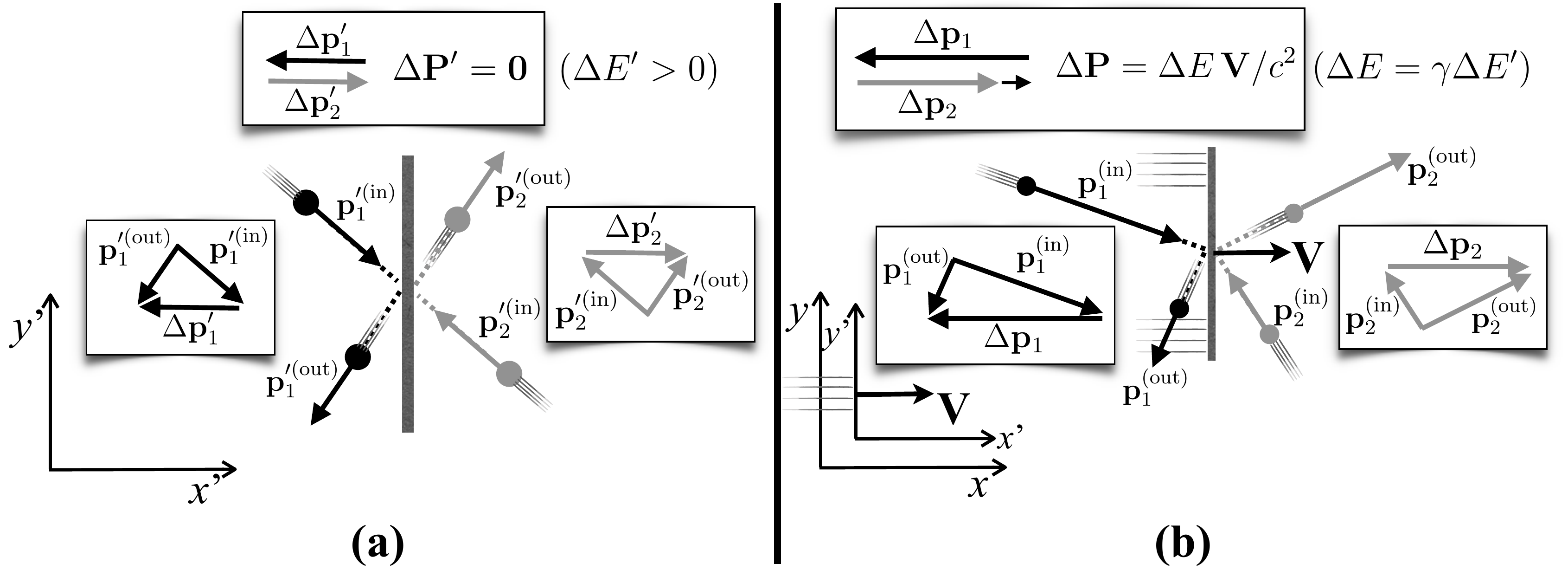}
\caption{(a) Symmetric inelastic collisions of two identical particles with a surface at rest. The symmetry of the setup
leads to no net momentum transfer to the surface. (b) The same process analyzed from
an inertial frame w.r.t.\ which the surface is moving with velocity ${\bf V}$. In this new frame, the collisions are no longer
symmetric and a net momentum $\Delta{\bf P} = \Delta E {\bf V}/c^2$ is transferred 
to the surface, where $\Delta E$ is the
energy delivered into the surface during the process. Assuming processes like this occurring at a constant rate,
a net force given by ${\bf F} = W {\bf V}/c^2$ would be exerted on the surface, where $W$ is the 
rate at which energy is delivered into the surface. By modeling a thermal reservoir as an isotropic  bath 
of particles, this implies that an object at rest in a thermal reservoir is subject to a net force ${\bf F} = W {\bf V}/c^2$ 
when seen from a reference frame w.r.t.\ which the whole system (object and reservoir) is moving --- with $W$ being 
the rate at which energy (heat) is absorbed by the object. {Note that  this force is exactly the one needed
to keep an
object with increasing mass, $dM/dt = W/c^2$, in uniform motion.}
}
\label{fig:model}
\end{figure}

Consider the symmetric process depicted in Fig.~\ref{fig:model}(a), where two identical particles with opposite momenta
(${\bf p}^\prime_2 = -{\bf p}^\prime_1$) and vanishing total angular momentum
collide with a surface at rest. We shall allow the collisions to be inelastic, each delivering an energy 
$\Delta E^\prime/2$ into
the surface. The symmetry of the setup makes it clear that, in this frame, the net 
momentum transferred  to the surface is zero ($\Delta {\bf P}^\prime =- (\Delta{\bf p}^\prime_1+\Delta
{\bf p}^\prime_2) = {\bf 0}$). In Fig.~\ref{fig:model}(b), the {\it same} process is depicted as seen from an inertial frame
w.r.t.\ which the surface moves with velocity ${\bf V}$. Obviously, the whole process is determined from its description
above; all one has to do is to Lorentz transform the primed quantities to this new frame. By doing so --- which is
a textbook exercise ---, one realizes that
the momentum exchanges between the particles and the surface are no longer symmetric ($\Delta {\bf p}_2\neq - \Delta
{\bf p}_1$) and that a net momentum  $\Delta {\bf P} = \Delta E {\bf V}/c^2$ is transferred to the surface, where 
$\Delta E = \gamma \Delta E^\prime$ is the net energy delivered into the surface in this frame 
($\gamma$ is the Lorentz factor). The implication is clear: 
 inelastic collisions which are symmetric in the rest frame
of the surface exert a {\it net force} on the surface when analyzed from  inertial frames w.r.t.\ which the surface is 
moving.

\begin{figure}
\includegraphics[scale=0.34]{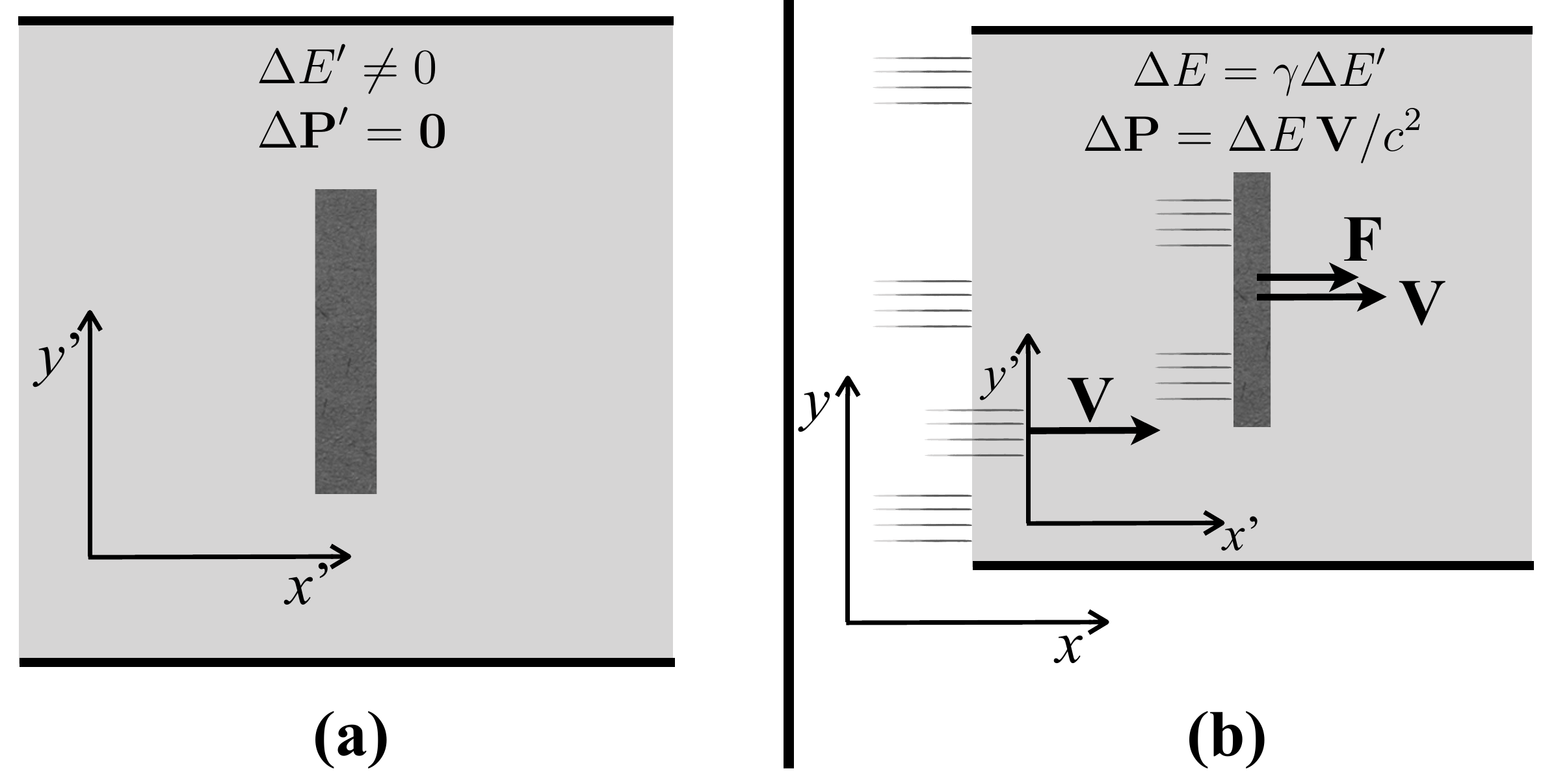}
\caption{Force applied on a moving object by a comoving thermal reservoir.
(a)~An object at rest in the reservoir's frame  
exchanges an amount $\Delta E'$ of energy in a time interval $\Delta t'$, 
with no  momentum transfer (due to the
symmetry of the reservoir in its rest frame). 
According to mass-energy equivalence, this corresponds to a (rest-)mass variation
$\Delta M'= \Delta E'/c^2$.
(b)~The same situation seen from another reference frame: a variation  $\Delta M = \gamma \Delta M'$ in the object's
mass, at constant velocity ${\bf V}$, corresponds to
a momentum transfer $\Delta {\bf P} = \Delta M {\bf V}= \gamma \Delta E'\,{\bf V}/c^2$ from the reservoir, in
a time interval $\Delta t = \gamma \Delta t'$. Therefore, in this frame, the reservoir must (and does) apply a net
force ${\bf F}=\Delta {\bf P}/\Delta t = \Delta E'\,{\bf V}/(c^2\Delta t') = W{\bf V}/c^2$ on the object, where $W=\Delta E'/\Delta t'=\Delta E/\Delta t$ 
is the
energy exchange rate. Causality/locality ensures that this final result cannot depend  on
whether the energy exchange $\Delta E'$ is accumulated in the
object or if it is used to sustain a stationary heat flow, as in Fig.~\ref{fig:system}.}
\label{fig:force}
\end{figure}

Modeling a thermal reservoir as an isotropic bath of particles, the result above inevitably leads to the conclusion that an object
static (and symmetrically immersed~\footnote{If the immersion is not symmetric, there may appear an additional
contribution related to the fact that the net force may not be zero in the rest frame of the system.})
in a thermal reservoir is subject to a (purely relativistic) 
net force ${\bf F} = W{\bf V}/c^2$ when seen from an inertial frame
w.r.t.\ which the whole system (object and reservoir) is moving with velocity ${\bf V}$, with $W$ being the rate at which
energy (heat) is absorbed by the object. Although this may sound odd at first, it becomes quite obvious when 
one realizes the {\it need}  of an external force in order to keep the constant velocity of an object with increasing
rest energy (i.e., rest mass). In fact, the existence of this relativistic force ${\bf F} = W{\bf V}/c^2$ can be 
inferred from this more general argument, independent of microscopic modeling of the reservoir (see Fig.~\ref{fig:force}).
The importance of the microscopic collisional model is explicitly showing that the existence of
such a force does  {\it not} depend  on the fate of the absorbed energy $\Delta E$ --- for instance, whether it is accumulated in the object or  constantly drained to sustain a heat flow (as in 
Fig.~\ref{fig:system})~\footnote{The same conclusion can be reached in the more general argument presented in
Fig.~\ref{fig:force} by invoking
locality and causality: the force exerted by the reservoir at the boundary of the object when the energy is 
exchanged cannot depend
on the later use of this energy.}. {Therefore, although, strictly speaking, 
the situations depicted in Fig.~\ref{fig:system} and
Figs.~\ref{fig:model} and \ref{fig:force} represent different systems,
the origin of the forces seen according to the ``moving frame'' is the same in all of them.
For an interesting 
quantum microscopic scenario where this force is also needed to conciliate 
different inertial-frame descriptions, see Ref.~\cite{STB}.}

\end{document}